# A formal definition of loop unrolling with applications to test coverage

**Bertrand Meyer**

Constructor Institute of Technology & Eiffel Software

Bertrand.Meyer@inf.ethz.ch

## Abstract

Techniques to achieve various forms of test coverage, such as branch coverage, typically do not iterate loops; in other words, they treat a loop as a conditional, executed zero or one time. Existing work by the author and collaborators produces test suites guaranteeing full branch coverage. More recent work has shown that by "unrolling" loops the approach can find significantly more bugs. The present discussion provides the theoretical basis and precise definition for this concept of unrolling.

While initially motivated by the need to improve standard test coverage practices (which execute loop bodies only once), to better testing coverage, the framework presented here is applicable to any form of reasoning about loops.

*Version 2 (this version), 12 August 2024: improved the presentation and corrected mistakes. This version should still be considered a draft as the theorems stated (as "properties") have not been formally proved.*

*Version 1, 13 March 2024.*

## 1 General idea

The theoretical underpinnings for the present work are to be found in denotational semantics [1] (see also the author's introduction in [2]), abstract interpretation [3], and the author's "*Theory of Programs*" outline [4].

We consider a loop L written "**until** e **loop** B **end**". The informal semantics is the usual one: execute the instruction B ("body") 0 or more times, stopping as soon as e (the "exit condition") holds. Another common syntax is "**while** ¬ e **loop** B **end**" where ¬ e is the logical negation of e. (Note that the loop does not execute anything if e is true – equivalently, if ¬ e is false – at the start, unlike in the "**repeat**… **until**" variant which always executes B at least once, and can trivially be expressed in the framework used here.)

The goal of the present discussion is to study how we can – in particular for testing purposes -- approximate L by a sequence of nested conditionals:

$L_0$ = **check False end**           -- Inapplicable program

$L_{i+1}$ = **if** ¬ e **then** B; $L_i$ **end**

(Here an instruction **check** p **end**, where p is a Boolean property, has no effect if p has value true upon execution, and otherwise makes the entire program in which it appears inapplicable. One can think of it in this case as causing a run-time crash but more abstractly it is simply an incorrect instruction. In the special case used here, **check False end** defining $L_0$, is simply a program that is never applicable, regardless of the input. Subsequent $L_{i+1}$, for i ≥ 0, are defined for a growing set of possible inputs: those that, in each case, require at most i executions of B before rendering e **True**.)

Why is it important to produce such a sequence of approximations? A number of applications exist, for example in software verification and beyond, but the concrete impetus for the present study was to provide a path beyond the *path coverage* techniques used in program testing. Path coverage is supposed to mean that a test suite has covered all possible paths through the program, but for any significant program, which will include loops (or similar control mechanisms such as recursive calls) the set of possible execution paths is infinite. Going from one extreme to another, most practitioners accept a form of "branch" coverage which only requires two paths for every loop body: executed once, or not at all. It is desirable to use an intermediate solution between these two extremes: "unroll" the loop a variable number of times (not just one), tuning the unrolling level in accordance with the testing needs and the available computing resources (since a higher level requires more testing time).

This notion of unrolling requires a sound notation, as will be provided now.

# 2 Trace semantics and notations

For the purpose of this discussion, the semantics of a program P is given by the set Traces (P) of its (finite) traces for any given input. In fact we identify P with its traces.

## 2.1 Traces and states

A trace x is a finite, non-empty sequence of program states written $<x_1, x_2, …>$.

In this definition, a program state is defined by the values of the program variables (in a general sense, which for an OO program will include the whole heap) as well as a program location. Intuitively it corresponds to what you see if you stop the program during execution and look at what the debugger tells you.

If s is a state (an element in a trace) and e is an expression, s [e] is the value of e in state s. For example if s is the state resulting from executing the instruction sequence x := 2 ; y := 5 the value of s [x ∗ y] is 10.

The i-th element (state), of a trace x is, if defined, written $x_i$. Since traces are non-empty, there always exists a first state, $x_1$, and a last state, written $x_L$. A trace is "stationary" if it has only element, i.e. is of the form $<x_1>$. (In this case $x_1$ is the same as $x_L$.) A stationary trace corresponds to an empty execution, which leaves the state unchanged.

We write x + y = z (or z = x + y) to express that trace z is the concatenation of traces x and y in the following sense:

- $x_L = y_1$

- The sequence of states of z is the concatenation of the sequences for x and y with the duplication at the border ($x_L$ and $y_1$) removed, e.g. <m, n> + <n, o, p> = <m, n, o, p >.

This definition does not introduce an operator "+" but a property "z = x + y". Per the first condition above, we may only use concatenation, to satisfy this property, if the last element of the first trace, $x_L$, is the same as the first element of the second one, $y_1$.

We may use the "+" notation, subject to the same condition, for any number of concatenated traces, not just two.

If x += y = z:

- If x is stationary, then y = z; if y is stationary then x = z.
- We say that x is a **prefix** of z and write x <= z (the relation is a partial order). If y is not stationary then x is a ***proper*** prefix and we write x < z.
- We also say that z is an **extension** (resp. proper extension) of x. (y is a "suffix" of z but we will not need that notion.)

A **test** is a condition – in other words, a Boolean expression – on states.

A trace x **satisfies** a test v if s [v] holds for some state s in x. (Remember that a state includes a program location.)

Property: if x satisfies v and x <= z, then z satisfies v.

## 2.2 Trace sets

Instructions and programs (in the underlying programming language) will be defined by their **trace sets**. A trace set is what the name indicates: a set of traces.

**skip** is the set of stationary traces (those of the form <$x_1$>, with just one state).

**fail** is the empty trace set. (Note that traces themselves cannot be empty, as they always have an initial state and a final state – which are the same for a stationary trace – but a trace *set* can be empty.)

We say that a trace set A **tests** t, or that t is **a test of** A, if there is trace in A satisfying t.

If A and B are trace sets, A + B, also written A ; B, is the set of traces

$\{z \mid \exists x: A, y: B \mid z = x + y\}$

In other words, the set of all traces obtained by concatenating a trace from A and a trace from B.

Properties:

$(A \, ; \, \text{fail}) = (\text{fail} \, ; \, A) = \text{fail}$

$(A \, ; \, \text{skip}) = (\text{skip} \, ; \, A) = A$

The "$\leq$" and "<" operators between traces similarly extend to trace sets: $A \leq B$ is defined as

$\forall x: A \mid \exists y: B \mid x \leq y$    -- And similarly for "<".

We can "slice" trace sets by conditions, pre- and post-. If c is a Boolean expression and A is a set of traces:

$c \, / \, A \quad \triangleq \quad \{x: A \mid x_1 \, [c]\}$    -- *Restriction*: only retain traces whose initial state satisfies c

$A \setminus c \quad \triangleq \quad \{x: A \mid x_L \, [c] \}$    -- *Corestriction*: only retain traces whose initial state satisfies c

Since we define program semantics by traces, we may use the following notations for programs:

$A = B$         -- Holds if and only if Traces (A) = Traces (B) (trace equivalence).

$A \cup B$       -- The program P such that Traces (P) = Traces (A) $\cup$ Traces (B).

$A \subseteq B$       -- Holds if every trace of A is also a trace of B. Also: $\subset$ etc.

        $x \in A$     -- Trace x is a trace (a possible execution) of program A.

Properties[1]:

| | | |
|---|---|---|
| False / A | = | fail |
| True / A | = | A |
| A \ False | = | **fail** |
| A \ True | = | A |
| c / (d / A) | = | (c ∧ d) / A |
| (A \ c) \ d | = | A \ (c ∧ d) |
| (A \ c) ; ¬ c / B | = | **fail** |
| (A \ c) ; (d / B) | ⊆ | A ; B |
| (A \ c) ; (c / B) | = | (A \ c)_; B  =  A_; (c / B)     [COMBINE] |

    -- Proof: removing "/" (resp "\") does not affect the trace set,
    -- already filtered by the other operator.

| | | | |
|---|---|---|---|
| (v / A) ; B | = | v / (A ; B) | [MOVE1] |
| A ; (B \ v) | = | (A ; B) \ v | [MOVE2] |
| v / (A ∪ B) | = | (v / A) ∪ (v / B) | [DIST1] |
| (A ∪ B) \ v | = | (A \ v) ∪ (B \ v) | [DIST2] |
| A ; (B ∪ C) | = | (A; B) ∪ (A ; C) | [DIST3] |
| (A ; B) ; C | = | (A ; C) ∪ (B ; C ) | [DIST3] |

If t is a test of A and A ≤ B, then t is also a test of B.

We will now use these concepts to define programming constructs and what they test.

## 3. Defining control structures

The standard program instructions and control structures are easy to express as trace sets.

We have already seen **skip** (defined as the set of one-element Traces) and **fail** (defined as the empty trace set).

Sequencing (block structure) has also been defined already through the operator "+" or its equivalent ";" which corresponds to the use of this symbol of programming languages. Here it correspondingly concatenates traces.

### 3.1. Power

Although "power" is not a common construct in programming languages (where it is subsumed by loops) it is convenient to define it in terms of trace sets. Inductively

---

[1] Currently these properties, and others asserted below, have been cursorily checked but not formally proved.

$$A^0 \triangleq \text{skip}$$
$$A^{i+1} \triangleq (A \,;\, A^i).$$

It corresponds to iterating A i times.

### 3.2. Conditionals

We define the conditional instruction "**if** v **then** A **end**" as

$$C \triangleq (\neg v \,/\, \textbf{skip}) \cup (v \,/\, A) \qquad \text{[COND]}$$

(For the present discussion we will not need the commonly used version of the conditional instruction including an **else** part, but it is trivial to add.)

The definition corresponds to the intuitive semantics of conditionals: an execution of C does nothing if v has value False, and otherwise is an execution of A.

### 3.3. Defining loops

We define the loop instruction L, written in programming language notation in the form

> **until** e **loop** B **end**

as

$$L \triangleq \bigcup_{i:\,\mathbb{N}} (\neg e \,/\, B)^i \setminus e \qquad \text{[LOOP\_1]}$$

This definition corresponds to the intuitive semantics of loops: an execution of L consists of 0 or more executions of B, from states in which e does not hold, such that the *last* of them produces a state where e holds.

Since the definition is a union, we can equivalently replace each element by the union of the preceding ones:

$$L = \bigcup L_i \qquad \text{[LOOP\_2]}$$

where

$$L_i \triangleq \bigcup_{j\,<\,i} (\neg e \,/\, B)^j \setminus e \qquad \text{[DEF\_ } L_i \text{]}$$

$L_i$ describes executions that achieve e by executing B repeatedly, but (strictly) less than i times. In particular, $L_0$ is an empty set, meaning **fail**, and

$$L_1 = \textbf{skip} \setminus e = e \,/\, \textbf{skip} \qquad \text{[LOOP\_SKIP1]}$$

We can express the $L_i$ sequence (the sequence whose union of all terms defines L) inductively as

$$L_0 \triangleq \textbf{fail}$$
$$L_{i+1} \triangleq L_i \cup ((\neg e \,/\, B)^i \setminus e)$$

## 4. A loop as a recursive conditional

One way to look at the loop L = "**until** e **loop** B **end**" is as a solution to the fixpoint equation

$$L = \text{if } \neg e \text{ then } B; L \text{ end} \qquad [\text{FIX}]$$

Rather than proving directly that loops as defined above (through the sequence $L_i$) satisfy [FIX], we consider the following sequence of programs inspired by this equation:

$$\underline{L}_0 \triangleq \textbf{fail}$$

$$\underline{L}_{i+1} \triangleq \text{if } \neg e \text{ then } B; \underline{L}_i \text{ end}$$

$$= (e\,/\,\textbf{skip}) \cup \neg e\,/\,(B; \underline{L}_i) \qquad \text{-- by the definition of conditionals}$$

It yields a proposed alternative definition $\underline{L}$ for loops:

$$\underline{L} \triangleq \bigcup_{i:\mathbb{N}} \underline{L}_i$$

## 4.1. The two views are equivalent

We will now prove that the definitions are equivalent, by showing by induction that $\underline{L}_i = L_i$ for all i. Both $L_0$ and $\underline{L}_0$ are **fail**. Then

$$L_{i+1} = \bigcup_{j \leq i} (\neg e\,/\,B)^j \setminus e \qquad \text{-- Definition of the L sequence [DEF\_ }L_i\text{]}$$

$$\underline{L}_{i+1} = (e\,/\,\textbf{skip}) \cup \neg e\,/\,(B; \underline{L}_i) \qquad \text{-- Definition of the }\underline{L}\text{ sequence}$$

$$= (e\,/\,\textbf{skip}) \cup (\neg e\,/\,(B; L_i)) \qquad \text{-- Induction hypothesis}$$

$$= L_1 \cup (\neg e\,/\,(B; L_i)) \qquad \text{-- From [LOOP\_SKIP1]}$$

$$= L_1 \cup (\neg e\,/\,(B; \bigcup_{j < i} ((\neg e\,/\,B)^j \setminus e)) \qquad \text{-- Definition of the L sequence}$$

$$= L_1 \cup ((\neg e\,/\,B); \bigcup_{j < i} ((\neg e\,/\,B)^j \setminus e)) \qquad \text{-- From [MOVE1]}$$

$$= L_1 \cup \bigcup_{j < i} (\neg e\,/\,B); ((\neg e\,/\,B)^j \setminus e) \qquad \text{-- from [DIST3]}$$

$$= L_1 \cup \bigcup_{j < i} ((\neg e\,/\,B); (\neg e\,/\,B)^j) \setminus e \qquad \text{-- from [MOV2]}$$

$$= L_1 \cup \bigcup_{j < i} ((\neg e\,/\,B)^{j+1}) \setminus e \qquad \text{-- Definition of power operator}$$

$$= L_1 \cup \bigcup_{j: 1..i} ((\neg e\,/\,B)^j) \setminus e \qquad \text{-- Change of index}$$

$$= L_1 \cup (L_i - L_1) \qquad \text{-- Definition of }L_i\text{ again}$$

$$\qquad \text{-- "}-\text{" is set difference}$$

$$= L_i \qquad \text{-- QED}$$

Property: $L_i \subseteq L$ for every i. (This is also true of $\underline{L}_i$ since it is the same as $L_i$.)

### 4.2. Some consequences

We call $L_i$ the **i-unrolling** of the loop L. It is of the form

$L_i \triangleq$     **if not** e **then**
        B
        **if not** e **then**
            B
            **if not** e **then**
                …
                **if not** e **then**
                    B
                    **check False end**     -- Corresponds to **fail**
                **end**
            …
            **end**
        **end**
    **end**

with exactly i occurrences of B (i.e. if i = 0 the instruction fails, if i = 1 it executes B once or fails, if i = 2 it executes B once or twice or fails etc.).

In the general case an i-unrolling executes B at most i times if it can do so with e each time not satisfied, and otherwise fails.

For every trace x of L, there is a smallest i such that x is a trace of $L_i$. By the definitions, x is also a trace of $L_j$ for all j > i. (Recall that $L_i \subseteq L_j$.)

As a consequence, for every test t of L, there is a minimum i such that t is a test of the i-th unrolling (and all subsequent ones).

This development gives us the theoretical framework that we need to unroll loops in our testing strategy. The default unrolling level is 1 (we treat a loop like an **if** … **then** … **end**). The more we unroll, the more extensive the tests will be.

A "bug" is a test for a specific condition (an incorrectness condition). Note that since an i+1-unrolling includes all the traces, and hence all the bugs, of an i-unrolling, the number of bugs found by a test can only be an increasing function of the unrolling level. (Otherwise, there is something wrong with the implementation of the strategy.)